\documentclass[conference]{IEEEtran}
\IEEEoverridecommandlockouts
\usepackage{cite}
\usepackage{amsmath,amssymb,amsfonts}
\usepackage{algorithmic}
\usepackage{graphicx}
\usepackage{textcomp}
\usepackage{xcolor}
\def\BibTeX{{\rm B\kern-.05em{\sc i\kern-.025em b}\kern-.08em
    T\kern-.1667em\lower.7ex\hbox{E}\kern-.125emX}}
        
\usepackage{fancyhdr}
\usepackage{comment}
\usepackage[caption=false]{subfig}
\usepackage{url}

\begin{document}

\title{From Pixels to Affect: \\A Study on Games and Player Experience
\thanks{This paper is funded, in part, by the H2020 project Com N Play Science (project no: 787476).}
}

\author{
\IEEEauthorblockN{Konstantinos Makantasis}
\textit{Institute of Digital Games}
 \\University of Malta,
 \\konstantinos.makantasis@um.edu.mt
 \and
 \IEEEauthorblockN{Antonios Liapis}
\textit{Institute of Digital Games}
 \\University of Malta,
 \\antonios.liapis@um.edu.mt
 \and
 \IEEEauthorblockN{Georgios N. Yannakakis}
\textit{Institute of Digital Games}
 \\University of Malta,
 \\georgios.yannakakis@um.edu.mt
}

\maketitle
\thispagestyle{fancy}

\begin{abstract}
Is it possible to predict the affect of a user just by observing her behavioral interaction through a video? How can we, for instance, predict a user's arousal in games by merely looking at the screen during play? In this paper we address these questions by employing three dissimilar deep convolutional neural network architectures in our attempt to learn the underlying mapping between video streams of gameplay and the player's arousal. We test the algorithms in an annotated dataset of 50 gameplay videos of a survival shooter game and evaluate the deep learned models' capacity to classify high vs low arousal levels. Our key findings with the demanding leave-one-video-out validation method reveal accuracies of over 78\% on average and 98\% at best. While this study focuses on games and player experience as a test domain, the findings and methodology are directly relevant to any affective computing area, introducing a general and user-agnostic approach for modeling affect.
\end{abstract}


\section{Introduction}\label{sec:introduction}

Designing general methods that are capable of performing equally well across various tasks has been a traditional vision of artificial intelligence \cite{goertzel2007artificial}. A milestone study in that direction is the work of Mnih \emph{et al.} \cite{mnih2015human} who achieved superhuman performance when playing several 2D games by merely observing the pixels of the screen. As impressive as these results might be, they are still limited to a particular set of tasks an agent needs to perform (i.e. play 2D Atari games) with clearly-defined objectives (i.e. maximize score). To which degree, however, could such general pixel-based representations learn to predict subjectively-defined notions such as emotion?

In this paper we attempt to address the above question based on the assumption that the behavior captured via the video of an interaction interweaves aspects of user experience that computer vision algorithms may detect. 
Thus, our key hypothesis is that we can construct accurate models of affect based only on the pixels of the interaction. In the current study we test this hypothesis in the domain of games
by assuming that there is an unknown underlying function between what a player sees on the screen during a gameplay session and the level of arousal in the game. We use games as our initial domain in this endeavor, as gameplay videos have the unique property of overlaying the game context onto aspects of playing behavior and affect. Given that player affect is already embedded in the context of playing, the dominant affective computing practice suggesting the fusion of context with affect is not necessary in this domain \cite{mcduff2014internet,ringer2018deep,zeng2009survey,mcduff2010affect,zafeiriou2017deep,wollmer2013youtube}.
Our approach is general and applicable to a variety of interaction domains beyond games since it only relies on decontextualized input (i.e. raw pixel values).

Given the spatio-temporal nature of the task, we use three types of deep convolutional neural network (CNN) architectures to classify between low and high values of annotated arousal traces based on a video frame or a video sequence.
In particular, we test the CNNs in a dataset of 50 gameplay videos of a 3D survival shooter game. All videos have been annotated for arousal by the players themselves (first-person annotation) using the \emph{RankTrace} \cite{lopes2017ranktrace} continuous annotation tool. Our key findings suggest that the task of predicting affect from the pixels of the experienced content is not only possible but also very accurate. Specifically, the obtained models of arousal are able to achieve average accuracies of over 78\% using the demanding leave-one-video-out cross-validation method; the best models we obtained yield accuracies higher than 98\%. The results also demonstrate---at least for the examined game---that player experience can be captured solely through on-screen pixels in a highly accurate and general fashion.

This paper is novel in several ways. First, this is the first attempt to model player affect just by observing the context of the interaction and not through any other direct manifestation of emotion or modality of user input; in that regard the solution we offer is \emph{general} and \emph{user-agnostic}. Second, to the best of our knowledge, this is the first time a study attempts to map directly from gameplay screen to game experience and infer a function between the two. Finally, three CNNs variants are compared for their ability to infer such a mapping in affective computing; the high accuracy values obtained demonstrate their suitability for the task.

\section{Related Work}\label{sec:related_work}
This section covers the related areas of affect modeling via videos, deep learning for images and videos, and affect modeling in games.

\subsection{Video-Based Affect Modeling}\label{sec:relatedwork_affectivecomputing}



Videos have been at the core of interest for both eliciting and modeling emotions in affective computing \cite{picard1995affective}. Typically, the video features a human face (or a group of faces) and emotion is modelled through the detection of facial cues (see \cite{littlewort2011cert,bartlett2006facial,mcduff2014internet} among many) due to theoretical frameworks and evidence supporting that facial expressions can convey emotion \cite{ambadar2005enigmatic,bassili1979emotion,ekman1992argument}. Beyond the facial expression of a subject, aspects such as the body posture \cite{kleinsmith2007posture,kleinsmith2011posturedb}, gestures \cite{glowinski2008gesture} or gait \cite{montepare1987gait,li2016emotion}, have been used as input for modeling affect.

To estimate the affective responses elicited to a person by external stimuli, affect annotations of such responses are required naturally. 
Indicatively, Chen \emph{et al.} \cite{chen2017gif} created a database of GIF animations, which users could rank across several affective dimensions, and modelled affect based on visual and tag features of the GIFs. In general, the onerous task of annotation makes such tasks ``intrinsically a small-sample learning problem'' \cite{li2015cloud}. This makes data-intensive methods such as deep learning rather inappropriate.
%
However, recent advances in deep learning have spurred research interest in emotion expression corpora, with several medium- and large-scale datasets as surveyed in \cite{barros2018omg}. 
CNNs were first applied in \cite{baveye2015videoprediction} to predict dimensional affective scores from videos, but the issue of small samples (raised above) challenged CNN learning. In \cite{kollias2017recognition}, CNNs were combined with recurrent neural networks to model arousal-valence using the \textit{Aff-Wild} database \cite{zafeiriou2017aff}. In \cite{tzirakis2019real} the authors exploit deep end-to-end trainable networks for recognizing affect in real-world environments. Finally, McDuff \emph{et al.} \cite{mcduff2014internet} fused facial expression data and videos of advertisements to classify whether viewers liked the videos or were willing to view them again. 

The modeling work presented here is unconventional within the broader affective computing field as it utilizes videos as both the elicitor of emotion and the sole modality for modeling affect. In a sense, what we achieve with the proposed approach is a general method for modeling affect via videos, as neither facial nor bodily expression is available as input to the affect model. The obtained high accuracies---at least within the games domain--- suggest that this subject-agnostic perspective is not only possible but it also yields models of high predictive capacity.  

\subsection{Deep Learning for Images and Videos}\label{sec:relatedwork_deeplearning}

Conventional machine learning methods have often been used for pattern recognition in images, videos and other data types, but have been held back by the requirement that raw data needed to be transformed to a suitable representation via a handcrafted feature construction process based on expert knowledge. The recent success of \emph{deep learning} \cite{goodfellow2016deep} approaches is largely due to their ability to learn representations directly from the raw data via the composition of simple but nonlinear data transformations. Very complex functions can be learned by combining enough transformations, and deep learning has shown tremendous success in visual recognition \cite{krizhevsky2012imagenet}, natural language processing \cite{young2018recent} and agent control \cite{mnih2015human}.

Convolutional neural networks are deep learning models which apply two-dimensional trainable filters and pooling operations on the raw input, resulting in a hierarchy of increasingly complex features. By design, CNNs are able to encode the spatial information of their inputs. CNNs are therefore particularly powerful in discovering patterns in 2D images \cite{krizhevsky2012imagenet}. 
CNNs have also been applied for classification of video sequences, similar to this paper, using a frame-by-frame input or a 3D representation with a temporal dimension \cite{karpathy2014video,ji2013convolutional}. While Jia \emph{et al.} \cite{jia2014caffee} first used 3D CNNs on cropped parts of a video, Ji \emph{et al.} proposed 3D CNNs for any video classification under the assumption that ``2D ConvNets lose temporal information of the input signal right after every convolution operation'' \cite{ji2013convolutional}. The testbed of Ji \emph{et al.} was the C3D dataset ($1.1\cdot10^6$ videos of 487 sports categories) with video frames resized to $128\times171$ pixels. The seminal paper of Karpathy \emph{et al.} \cite{karpathy2014video} explored several architectures for fusing information over the temporal dimension, including an early fusion approach which combined RGB channels over time (as 4D CNN) and a late fusion which used two single-frame networks (each receiving frames spaced half a second apart) and compared outputs to derive global motion characteristics. Similarly to \cite{ji2013convolutional}, the work of Karpathy \emph{et al.} was also tested on the C3D dataset, but videos were resized and cropped to $170\times170$ pixels. 

This paper uses a game footage dataset which is far smaller than the data available to the above studies, which necessitated a simplification of both the video input (which was downscaled more aggressively and only used the brightness channel) and the CNN architecture (with far fewer trainable parameters).

\subsection{Affect Modeling in Games} 


Player modeling is the study of computational models of players, their behavioral patterns and  affective responses \cite{yannakakis2012playermodeling}. 
If target outputs are available, a player model considers some input modality regarding the player (e.g. their gameplay and physiology) and is trained to predict aspects of the in-game behavior or the player experience. 
Indicatively, in studies with \emph{Super Mario Bros.} (Nintendo, 1985) gameplay data (e.g.~number of deaths) combined with level features (e.g.~number of gaps) \cite{pedersen2009experience}, or the player's posture during gameplay \cite{shaker2013visual} were used to predict the player's reported affect.

This study advances the state of the art in player modelling by using solely raw gameplay information to model a player's emotions. Within the broader area of artificial intelligence and games \cite{yannakakis2018artificial}, the majority of the works that analyse and extract information from gameplay videos focus on inferring the strategy, structure and the physics of the games themselves \cite{guzdial2016deep,mnih2015human}. In this work, instead, we use the same kind of information for modelling a player's experience in a general fashion (from pixels to experience), ignoring the game per se. At the same time, the most common approaches for analysing player experience, besides game and gameplay information, heavily rely on direct measurements from players, such as face monitoring, speech and physiological signals; see e.g. \cite{ringer2018deep,shaker2013visual,martinez2013learning}). 
Unlike these approaches, our methodology relies solely of gameplay video information. This critical difference advances player experience modelling as the approach does not require access to intrusive player measurements collected under well-defined experimental settings, thus allowing the vast collection of data. As gameplay videos are already available over the web and produced daily in massive amounts, the approach is feasible and can potentially generalize to any game. 

\begin{figure*}[t]
 \begin{minipage}{1.0\linewidth}
 \centering
 \centerline{\fbox{\includegraphics[width=0.98\linewidth]{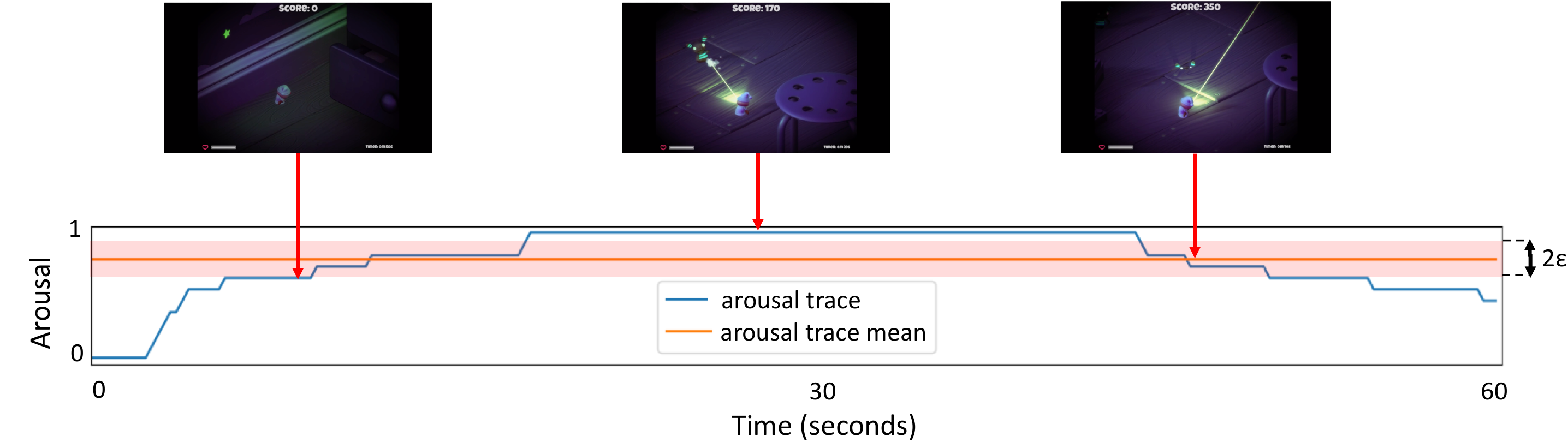}}}
 \end{minipage}
 \caption{The normalized to $[0,1]$ trace of affect (arousal) produced by RankTrace, the uncertainty zone define by $\epsilon$, and three indicative frames of one of the Survival Shooter gameplays.}
 \label{fig:annotations}
\end{figure*}

\section{Methodology}\label{sec:method}

This paper explores the degree to which frames and videos of gameplay footage can act as the sole predictors of a player's affective state. This section describes the gameplay dataset and how it was collected, 
the employed CNN architectures, as well as the dataset preparation process for training the CNNs.

\subsection{Dataset Description}\label{sec:method_dataset}

The gameplay videos we used in the experiments of this paper are captured from a shooter game developed in the \textit{Unity 3D} game engine. Specifically, we use the \textbf{Survival Shooter} \cite{camilleri2017generalmodels}, which is a game adapted from a tutorial package of Unity 3D. In this game the player has 60 seconds to shoot down as many hostile toys as possible and avoid running out of health due to toys colliding with the avatar. Hostile toys keep spawning at predetermined areas of the level and converge towards the player. The player's avatar has a gun that shoots bright laser beams, and can kill each toy with a few shots. Every toy killed adds to the player's score.

The data was collected from 25 different players who each produced and annotated two gameplay videos. Each player played a game session (60 seconds) and then annotated their recorded gameplay footage in terms of arousal. Annotation was carried out using the \textit{RankTrace} annotation tool \cite{lopes2017ranktrace} which allows the continuous and unbounded annotation of affect using the Griffin PowerMate wheel interface. Gameplay videos were captured at 30Hz (i.e.~30 frames per second) while the RankTrace tool provided four annotation samples per second. 
Figure \ref{fig:annotations} shows three indicative frames of the Survival Shooter gameplay and the annotations of arousal from RankTrace. 

The corpus of gameplay videos was cleaned by omitting gameplay footage under 15 seconds, resulting in a clean corpus of 45 gameplay videos and a total of $8,093$ annotations of arousal. While the average duration of playthroughs in this corpus is 44 seconds, in 60\% of the playthroughs the player survived for the full 60 seconds and completed the game level. 

\subsection{Training Data Preparation}\label{sec:method_processing}

In order to evaluate how CNNs can map raw video data to affective states, we train CNN models using as input individual frames that contain only spatial information, and video segments that contain both spatial and temporal information. This section describes the input and the output of the networks.

Since RankTrace provides unbounded annotations, we first convert the annotation values of each video to $[0,1]$ via min-max normalization and synchronize the recording frequency of videos (30Hz) with annotations (4Hz) by treating the arousal value of any frame without an annotation as the arousal value of the last annotated frame. In order to decrease the computational complexity of training and evaluating CNNs, we convert RGB video frames to grayscale and resize them to $72 \times 128$ pixels; this results in a more compact representation which considers only the brightness of the image and not its color. Due to the stark shadows and brightly lit avatar and projectiles in the Survivor Shooter, we consider that brightness is likely a core feature for extracting gameplay behavior. While RGB channels or a larger frame size could provide more information about the gameplay and affect dimensions, it would require substantially more data for CNNs to train on.


Regarding the input of the CNN, we have to decide which frames and video segments will be used as input points. Schindler and Van Gool \cite{schindler2008action} argue that a small number of subsequent frames are adequate to capture the content of a scene. Based on this argument, the authors of \cite{makantasis2016deep} achieve high human activity recognition rates by describing an activity with mini video batches of 8 subsequent frames. Motivated by these works, we also use 8 subsequent frames to characterize the player's state of affect. Specifically, the gameplay videos are split into non-overlapping segments of 8 subsequent frames which are used as input to the temporally aware CNN architectures. If the input is a single image, the last frame of each video segment is used.

The output of the CNN is straightforward to compute based on the 8-frame video segments. Since annotations are made at 4Hz, in most cases a video frame segment would include one annotation. In cases where two annotations are given within 8 frames, their average value is computed. RankTrace produces interval data and thus it may seem natural to state the problem as a regression task; given that we aim to offer a user-agnostic and general approach, however, we do not wish to make any assumptions regarding the value of the output as this may result in highly biased and user-specific models \cite{yannakakis2018ordinal}. For this reason we state our problem as a classification task and transform interval values into binary classes (low and high arousal) by using the mean value of each trace as the class splitting criterion (see Fig.~\ref{fig:annotations}). The class split may use an optional threshold parameter ($\epsilon$) to determine the zone within which arousal values around the mean are labelled as `uncertain' and ignored during classification. Detailed experiments with the $\epsilon$ parameter are conducted in Section \ref{sec:experiment_thresh}. While alternative ways of splitting the classes were considered (such as the area under the curve or the median), in this paper we include only experiments with the most intuitive way to split such a trace given its unbounded nature: its mean.

\subsection{CNN Architectures}

In this study we explore 
three different CNN architectures. The first two apply 2D trainable filters on the inputs (single frames or videos), while the third applies 3D trainable filters. All CNN architectures have the same number of convolutional and fully connected layers, the same number of filters at their corresponding convolutional layers and the same number of hidden neurons at their fully connected layer. This way we are able to fairly compare the skill of the three architectures to map video data to affective states, and at the same time to gain insights on the effect of temporal information to the classification task. It should be noted that current state-of-the-art CNNs for videos and images alike use much larger architectures (e.g.~\cite{karpathy2014video}); in this paper, however, we explore more compact architectures due to the small size of the dataset.

\subsubsection{2DFrameCNN}

\begin{figure}[!tb]
 \begin{minipage}{1.0\linewidth}
 \centering
 \centerline{\fbox{\includegraphics[width=0.98\linewidth]{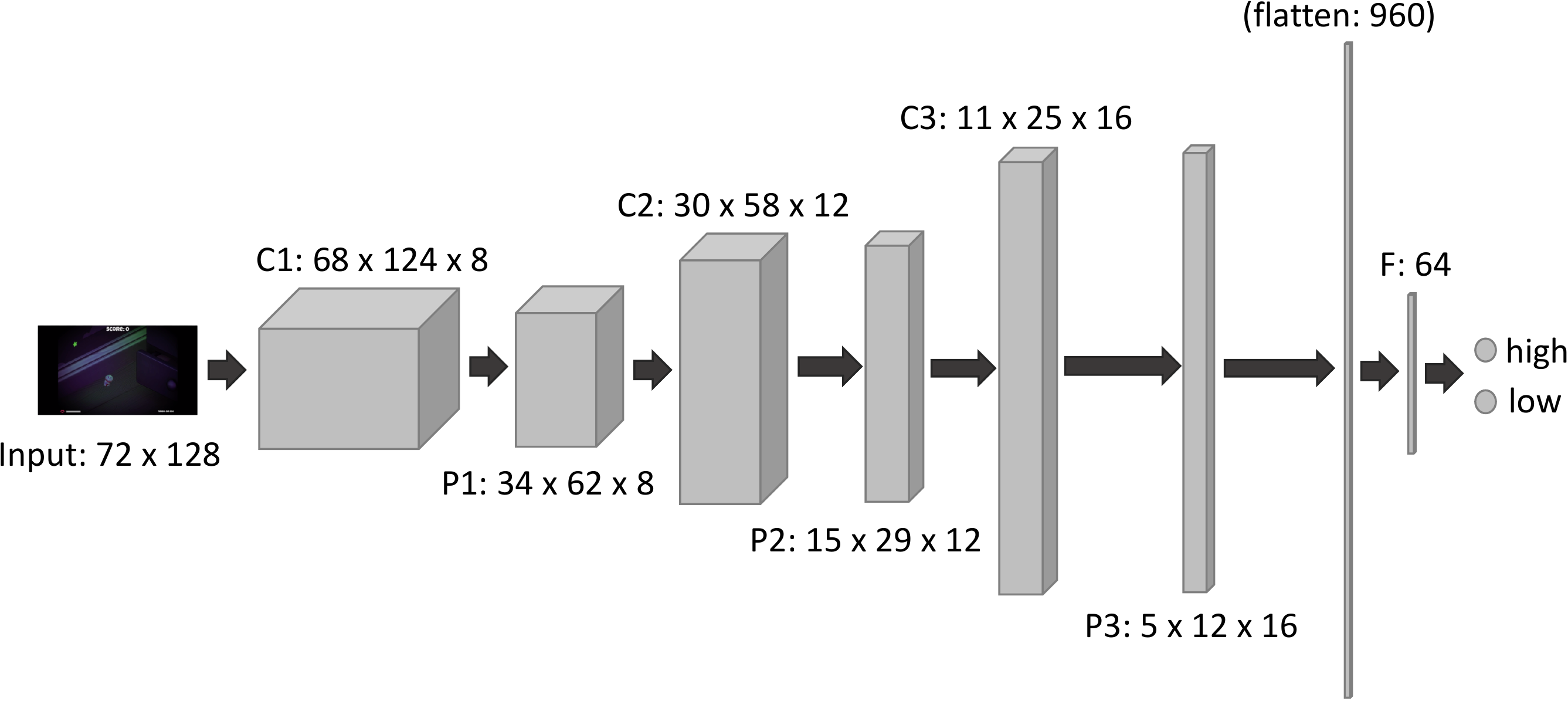}}}
 \end{minipage}
 \caption{The architecture of 2DFrameCNN. Convolutional layers are denoted with a ``C'', max pooling layers with a ``P'', and fully connected layers with an ``F''.}
 \label{fig:CNN}
\end{figure}

The first CNN architecture (see Fig. \ref{fig:CNN}) uses as input a single frame on which it applies 2D filters. The \emph{2DFrameCNN} architecture consists of three convolutional layers with 8, 12 and 16 filters, respectively, of size $5 \times 5$ pixels. Each convolutional layers is followed by a 2D max pooling layer of size $2 \times 2$. The output of convolutions is a feature vector of 960 elements, which is fed to a fully connected layer with 64 hidden neurons that connect to the output. This architecture has approximately $6.9\cdot10^4$ trainable parameters and exploits only the spatial information of the video data.

\subsubsection{2DSeqCNN}

The second CNN architecture applies 2D filters to input video segments. The \emph{2DSeqCNN} network has exactly the same topology as the 2DFrameCNN architecture but the number of trainable parameters is slightly higher (approximately $7\cdot10^4$) as the inputs are video sequences. 
This architecture implicitly exploits both the spatial and the temporal information of the data. 

\subsubsection{3DSeqCNN}

The third CNN architecture 
applies 3D filters to input video segments. As with the other architectures, \emph{3DSeqCNN} has three convolutional layers with 8, 12 and 16 filters, respectively, of size $5 \times 5 \times 2$ pixels. Each one of the convolutional layers is followed by a 3D max pooling layer of size $2 \times 2 \times 1$. The 3D convolutional layers produce a feature vector of 1,920 elements, which is fed to a fully connected layer with 64 neurons. Due to its 3D trainable filters, \emph{3DSeqCNN} has approximately $14.5\cdot10^4$ trainable parameters.
This architecture explicitly exploits both the spatial and the temporal information of the data due to the application of the trainable filter along the spatial and the temporal dimensions.

While 2DFrameCNN receives as input a single frame, both 2DSeqCNN and 3DSeqCNN receive as input a sequence of 8 frames, i.e. a time slice of the video lasting 267 milliseconds. In all three network architectures, we apply batch normalization on the features constructed by the convolutional layers before feeding them to the last fully connected layer, which in turn feeds two output neurons for binary classification. All of the hyperparameters of the CNN architectures are manually selected in an attempt to balance two different criteria: (a) computational complexity (training and evaluation times), and (b) learning complexity (ability to avoid under-/over-fitting). 

\section{Experiments}\label{sec:experiment}


To test our hypothesis that there is a learnable underlying function between affect and its visual manifestations on gameplay videos, in this section we use the three CNNs for classifying gameplay footage as \emph{high} or \emph{low} arousal (as discussed in Section \ref{sec:method_processing}). 
As mentioned earlier, this binary classification approach is well-suited for unbounded and continuous traces (as the mean of each annotation trace is different), and can produce a sufficiently rich dataset for deep learning. Section \ref{sec:experiment_nothresh} explores the performance of different CNN architectures on this naive split between high and low arousal, while Section \ref{sec:experiment_thresh} explores the impact of an uncertainty bound that filters out segments that are too close to the mean arousal value.

In all reported experiments, we follow the demanding \textit{leave-one-video-out} scheme \cite{mcduff2014internet}; this means that we use data from 44 videos to train the models and then we evaluate their performance on the data from the video that is not used for training (i.e. test set). This procedure is repeated 45 times until we test the performance of CNNs on the data from all videos. During the training of the models we also employ early stopping criteria to avoid overfitting. For early stopping, data of the 44 videos is shuffled and split further into a training set (90\% of the data) and a validation set for testing overfitting (10\% of the data). Early stopping is activated if the loss on the validation set does not improve for 15 training epochs. Reported accuracy is the classification accuracy on the test set, averaged from 45 runs. Significance is derived from the 95\% confidence interval of this test accuracy. The baseline accuracy is the average classification accuracy on the test set, when we always select the most common class in the 44 videos of the training set. Naturally, the baseline also indicates the distribution of the ground truth between the two classes. 


\subsection{Binary Classification of Arousal}\label{sec:experiment_nothresh}

The most straightforward way to classify segments of gameplay footage is based on the mean arousal value of the annotation trace, treating all annotations above the mean value as high arousal and below it as low arousal. This naive classification results to a total of $8,093$ data points (i.e. 8-frame segments assigned to a class) from all 45 videos. 

\begin{table}[!t]
    \centering
    \caption{Test accuracy for binary classification of different CNN architectures, and for different threshold values for classification ($\epsilon$). The 95\% confidence interval is included.}
    \label{table:all_accuracies}
    \begin{tabular}{l|r@{}l|r@{}l|r@{}l|r@{}l}
    \hline\hline
    $\epsilon$ & \multicolumn{2}{c|}{Baseline} & \multicolumn{2}{c|}{2DFrameCNN} & \multicolumn{2}{c|}{2DSeqCNN} & \multicolumn{2}{c}{3DSeqCNN}	\\
    \hline\hline
0.00 & 	51\% &$\pm$0.0\% &	70\% & $\pm$4.2\% &			74\% & $\pm$4.7\% &			73\% & $\pm$4.4\% \\
0.05 &56\% &$\pm$0.3\% &	72\% & $\pm$5.6\% &			73\% & $\pm$5\% &			73\% & $\pm$5.3\% \\
0.10 & 55\% &$\pm$0.3\% &	74\% & $\pm$5.7\% &			75\% & $\pm$5.6\% &			74\% & $\pm$5.7\% \\
0.20 & 50\% &$\pm$0.3\% &	77\% & $\pm$5.7\% &			78\% & $\pm$5.6\% &			77\% & $\pm$5.7\% \\
\hline
    \end{tabular}
\end{table}

The top row of Table \ref{table:all_accuracies} reports the average classification accuracy of the CNN models with the naive classification method ($\epsilon=0$). All models have accuracies over 20\% higher than the baseline classifier, which suggests that CNNs, regardless of the architecture used, have the capacity to map raw gameplay video to arousal binary states. The model that performs best is the 2DSeqCNN, which implicitly exploits the temporal information in the data. Its accuracy is over 3\% higher than the 2DFrameCNN which exploits only spatial information, but it is only slightly better than the 3DSeqCNN. The ability of the 3DSeqCNN to explicitly exploit the temporal information does not seem to significantly affect its performance. Comparing the performance of the 2DFrameCNN with the performances of the other two CNN models indicates that although the temporal information contributes to the learning process, the dominant information of the inputs comes from their spatial and not their temporal structure. This may be due to the very short duration of the input video segments (267 milliseconds), or due to strong predictors of arousal existent in the heads-up display of the game (see Section \ref{sec:experiment_activation}).

\subsection{Exploring the Uncertainty Bound of Arousal 
}\label{sec:experiment_thresh}

While classifying all data above the mean value of the arousal trace as high yields a large dataset, the somewhat arbitrary split of the dataset may misrepresent the underlying ground truth and also introduce split criterion biases \cite{martinez2014don,yannakakis2018ordinal}. Specifically, frames with arousal values around the mean would be classified as \emph{high} or \emph{low} based on trivial differences. To filter out annotations that are ambiguous (i.e. close to the man arousal value $\hat{A}$), we use the $\epsilon$ value and omit any datapoints with an arousal value $A$ within the \emph{uncertainty bound} determined by $\epsilon$: $\hat{A}-\epsilon < A < \hat{A}+\epsilon$ (see Fig. \ref{fig:annotations} for a graphical depiction). This section tests how the performance of the three CNN classifiers changes when three different threshold values $\epsilon=\{0.05,0.10,0.20\}$ remove ambiguous data points from the dataset. 

Table \ref{table:all_accuracies} shows the performance of different CNN architectures for different threshold values. It should be noted that removing datapoints affects the baseline values quite substantially as representatives of one class become more frequent than for the other class. Regardless, we see that the accuracy of all architectures increases when data with ambiguous arousal values is removed, especially for higher $\epsilon$ values. For $\epsilon=0.20$, the accuracy of all three CNN architectures is 26\% to 28\% higher than the baseline. 
The 2DFrameCNN also benefits from the cleaner dataset, being second in accuracy only to 2DSeqCNN for $\epsilon=0.10$ and $\epsilon=0.20$. The additional trainable parameters of 3DSeqCNN seem to require more data than what is available in the sparser datasets. Indeed, the number of total datapoints decreases by 12\% for $\epsilon=0.05$, by 25\% for $\epsilon=0.10$, and by 44\% for $\epsilon=0.20$ (for a total of $4,534$ datapoints). It is obvious that having a cleaner but more compact dataset can allow the less complex architectures (2DFrameCNN, 2DSeqCNN) to derive more accurate models but can challenge complex architectures (3DSeqCNN). The trade-off poses an interesting problem moving forward for similar tasks of gameplay annotation.

\subsection{Analysis of Findings}\label{sec:experiment_activation}

Experiments showed that it is possible to produce surprisingly accurate models of players' arousal from on-screen gameplay footage alone---even from a single frame snapshot. Especially when removing data with ambiguous arousal annotations, a model of 2DFrameCNN can reach a test accuracy of 98\% (at $\epsilon=0.20$), although on average the test accuracy is at 77\%. It is more interesting, however, to observe which features of the screen differentiate frames or videos into low-arousal or high-arousal classes. This can be achieved by showing which parts of the frame have the most influence on the model's prediction, e.g. via Gradient-weighted Class Activation Mapping \cite{selvaraju2017grad}. This method computes the gradient of an output node with respect to the nodes of a convolutional layer, given a particular input. By multiplying the input with the gradient, averaging over all nodes in the layer and normalizing the resulting values, we obtain a heatmap that shows how much each area of the input contributed to increasing the value of the output node.

\begin{figure}[t]
\centering
\subfloat[Frame of gameplay footage (in full resolution and full-color)]{\includegraphics[width=0.95\linewidth]{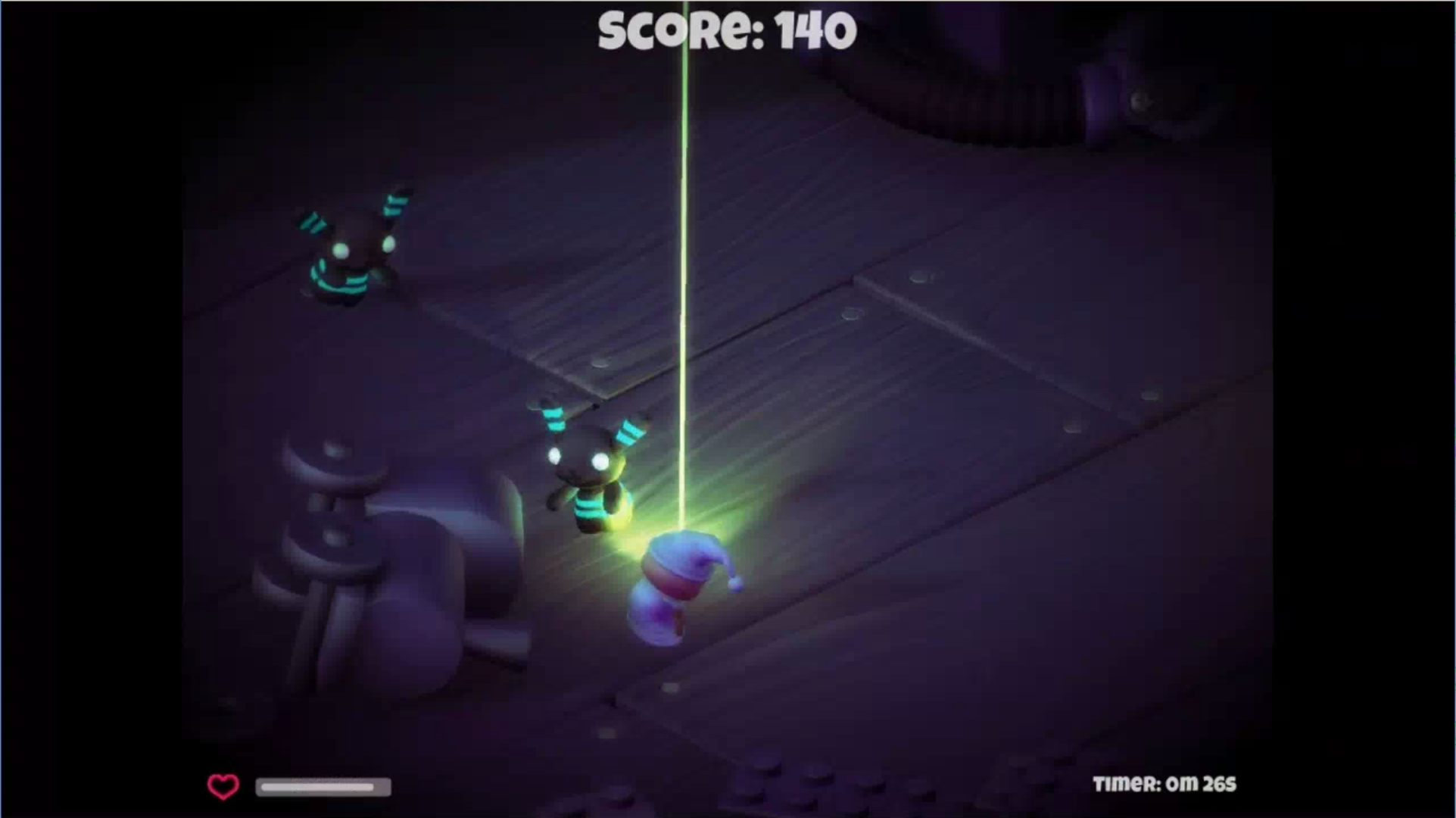}\label{fig:activationmaps_base}}\\
\subfloat[Activation of \textbf{Low} Arousal]{\includegraphics[width=0.45\linewidth]{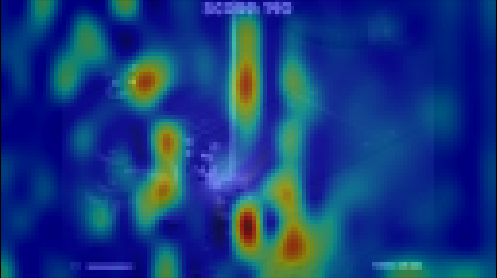}\label{fig:activationmaps_1}}\quad
\subfloat[Activation of \textbf{High} Arousal]{\includegraphics[width=0.45\linewidth]{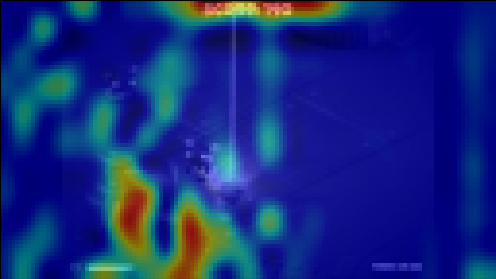}\label{fig:activationmaps_2}}
\caption{Activation maps for a sample frame of the game}
\label{fig:activationmaps}
\end{figure}

Figure~\ref{fig:activationmaps} shows the activation maps for low versus high arousal of a sample gameplay frame, calculated based on the 2DFrameCNN. While 2DSeqCNN has higher accuracies, it is far more challenging to visually capture the sequence on paper so we opt for the frame-only information of 2DFrameCNN. We immediately observe that both low and high arousal predictors focus on aspects of the heads-up display (HUD) which are overlaid on the 3D world where the player navigates, shoots and collides with hostile toys. 
Specifically, the score at the top center of the screen contributes substantially to high arousal. Interestingly, the score keeps increasing during the progression of the game as the player kills more and more hostile toys. The impact of time passed in the game---and by extent increasing score---on arousal can be corroborated by the annotations themselves: in most cases the annotators kept increasing the arousal level as time went by rather than decreasing it. Tellingly, of all arousal value changes in the entire dataset, 807 instances were increases and 297 were decreases. Thus, both score and time remaining would be simple indicators of low or high arousal. Interestingly, the HUD element of the player's health was not considered for either class. Among other features of the 3D gameworld, hostile toys are captured by the low arousal output, while an obstacle next to the player is captured by the high arousal output. It is less clear what other areas activated on the screens of Fig.~\ref{fig:activationmaps} capture with regards to arousal.

\section{Discussion}\label{sec:discussion}

This paper presented the first attempt, to our knowledge, of modelling affect solely via videos that do not display human behavior \emph{directly}; such videos of interaction instead display human behavior in an \emph{indirect} manner as emotion is manifested through and annotated on the video per se. We also introduced the first modeling attempt of players' affective states based on the on-screen captured gameplay alone. Using a time window of 8 frames and selecting either a single frame or the frame sequence within that time window, a number of CNN architectures were tested. Results show that processing the gameplay footage as short videos results in higher classification accuracy, although in general all three models perform comparably. Moreover, when data within an uncertainty bound around the trace's mean arousal are not considered, the smaller remaining dataset challenges 3D convolutional layers but yields highly accurate models (approximately 78\% accuracy, on average) for simpler networks based on frames or frame-by-frame processing of videos. Despite this paper being a first attempt at a challenging task of predicting player affect from gameplay pixels, the results are promising and point to a number of extensions in future work. We discuss these below.

As this was an initial exploratory study, there is a number of assumptions made for both the input and the output of the affect model. In terms of input, we used only the brightness channel of the gameplay footage; in part, this was because of the structure of the game itself, due to the high-contrast ``horror'' aesthetic, and because one channel allowed us to train models faster and with the few data points at hand. Future experiments, however, should explore other formats for CNN-based video classification in the literature, such as scaling the input to be a higher-resolution image, and using hand-crafted channels that include edge detection \cite{ji2013convolutional} or RGB channels \cite{karpathy2014video}. While the goal of this study was to detect player arousal from gameplay footage alone, future studies could explore how fusing gameplay footage with other information streams such as gameplay logs and physiological data \cite{camilleri2017generalmodels} would affect the model's accuracy. In terms of the affect labels (output), taking the mean arousal value (normalized to a player's full trace) within a time window was an intuitive solution, but could be expanded on and refined. 
More relative ways of processing annotations within a time window, such as amplitude and average gradient \cite{camilleri2017generalmodels,lopes2017ranktrace}, could be explored. Moreover, the video information (processed through a CNN) could be used to predict not the class of high or low arousal but instead whether there is an increase or decrease from the previous time window. This method would better align with the temporal nature of gameplay videos, but would likely decrease the size of the dataset as many subsequent time windows have the same mean arousal value (see Fig.~\ref{fig:annotations}). Finally, using a sliding time window rather than a non-overlapping window of 8 frames would increase the size of the dataset and perhaps better capture all annotations. 

\section{Conclusion}\label{sec:conclusion}

In this paper we introduced a general method that captures affect solely from videos which embed forms of human computer interaction but without humans explicitly depicted in the video. 
Using games as our domain, 
we explored how gameplay footage can be processed and fed to three different convolutional neural network architectures that, in turn, predict a player's arousal levels in a binary fashion. The obtained models of arousal trained this way yield accuracies of up to 78\% on average (98\% at best). Our analysis also reveals the different on-screen aspects that contribute to higher vs. lower arousal in the testbed game.
While this initial study focuses on games as a domain, the findings and methodology are directly relevant to any affective computing area, introducing a general and user-agnostic approach for modeling affect.




\begin{thebibliography}{10}
	
	\bibitem{goertzel2007artificial}
	Ben Goertzel and Cassio Pennachin,
	\newblock {\em Artificial general intelligence}, vol.~2,
	\newblock Springer, 2007.
	
	\bibitem{mnih2015human}
	Volodymyr Mnih, Koray Kavukcuoglu, David Silver, Andrei~A Rusu, Joel Veness,
	Marc~G Bellemare, Alex Graves, Martin Riedmiller, Andreas~K Fidjeland, Georg
	Ostrovski, et~al.,
	\newblock ``Human-level control through deep reinforcement learning,''
	\newblock {\em Nature}, vol. 518, no. 7540, 2015.
	
	\bibitem{mcduff2014internet}
	Daniel McDuff, Rana el~Kaliouby, David Demirdjian, and Rosalind Picard,
	\newblock ``Predicting online media effectiveness based on smile
	responsesgathered over the internet,''
	\newblock in {\em Image and Vision Computing}. Elsevier, 2014.
	
	\bibitem{ringer2018deep}
	Charles Ringer and Mihalis~A Nicolaou,
	\newblock ``Deep unsupervised multi-view detection of video game stream
	highlights,''
	\newblock in {\em Proceedings of {FDG}}, 2018.
	
	\bibitem{zeng2009survey}
	Zhihong Zeng, Maja Pantic, Glenn~I Roisman, and Thomas~S Huang,
	\newblock ``A survey of affect recognition methods: Audio, visual, and
	spontaneous expressions,''
	\newblock {\em IEEE Trans. on Pattern Analysis and Machine Intelligence}, vol.
	31, no. 1, pp. 39--58, 2009.
	
	\bibitem{mcduff2010affect}
	Daniel McDuff, Rana El~Kaliouby, Karim Kassam, and Rosalind Picard,
	\newblock ``Affect valence inference from facial action unit spectrograms,''
	\newblock in {\em Proc. of the IEEE Computer Society Conf. on Computer Vision
		and Pattern Recognition Workshops}. IEEE, 2010, pp. 17--24.
	
	\bibitem{zafeiriou2017deep}
	Lazaros Zafeiriou, Stefanos Zafeiriou, and Maja Pantic,
	\newblock ``Deep analysis of facial behavioral dynamics,''
	\newblock in {\em Proc. of the IEEE Conf. on Computer Vision and Pattern
		Recognition Workshops}, 2017.
	
	\bibitem{wollmer2013youtube}
	Martin W{\"o}llmer, Felix Weninger, Tobias Knaup, Bj{\"o}rn Schuller, Congkai
	Sun, Kenji Sagae, and Louis-Philippe Morency,
	\newblock ``Youtube movie reviews: Sentiment analysis in an audio-visual
	context,''
	\newblock {\em IEEE Intelligent Systems}, vol. 28, no. 3, pp. 46--53, 2013.
	
	\bibitem{lopes2017ranktrace}
	Phil Lopes, Georgios~N Yannakakis, and Antonios Liapis,
	\newblock ``Ranktrace: Relative and unbounded affect annotation,''
	\newblock in {\em Proc. of the Intl. Conf. on Affective Computing and
		Intelligent Interaction}, 2017, pp. 158--163.
	
	\bibitem{picard1995affective}
	Rosalind Picard,
	\newblock ``Affective computing,''
	\newblock Tech. {R}ep., MIT, 1995.
	
	\bibitem{littlewort2011cert}
	G.~Littlewort, J.~Whitehill, T.~Wu, I.~Fasel, M.~Frank, J.~Movellan, and
	M.~Bartlett,
	\newblock ``{The computer expression recognition toolbox (CERT)},''
	\newblock in {\em Proc. of Face and Gesture}, 2011.
	
	\bibitem{bartlett2006facial}
	M.~Bartlett, G.~Littlewort, M.~Frank, C.~Lainscsek, I.~Fasel, and J.~Movellan,
	\newblock ``Automatic recognition of facial actions in spontaneous
	expressions,''
	\newblock {\em Journal of Multimedia}, vol. 1, no. 6, 2006.
	
	\bibitem{ambadar2005enigmatic}
	Z.~Ambadar, J.~Schooler, and J.~Cohn,
	\newblock ``Deciphering the enigmatic face. the importance of facial dynamics
	in interpreting subtle facial expressions,''
	\newblock {\em Psychological Science}, vol. 16, 2005.
	
	\bibitem{bassili1979emotion}
	J.~Bassili,
	\newblock ``Emotion recognition: The role of facial movement and the relative
	importance of upper and lower areas of the face,''
	\newblock {\em Journal of personality and social psychology}, vol. 37, no. 11,
	1979.
	
	\bibitem{ekman1992argument}
	Paul Ekman,
	\newblock ``An argument for basic emotions,''
	\newblock {\em Cognition \& emotion}, vol. 6, no. 3-4, pp. 169--200, 1992.
	
	\bibitem{kleinsmith2007posture}
	Andrea Kleinsmith and Nadia Bianchi-Berthouze,
	\newblock ``Recognizing affective dimensions from body posture,''
	\newblock in {\em Proc. of the Intl. Conf. on Affective Computing and
		Intelligent Interaction}, 2007.
	
	\bibitem{kleinsmith2011posturedb}
	Bianchi-berthouze N. Steed~A. Kleinsmith, A.,
	\newblock ``Automatic recognition of non-acted affective postures,''
	\newblock {\em IEEE Trans. on Systems, Man and Cybernetics}, 2011.
	
	\bibitem{glowinski2008gesture}
	D.~{Glowinski}, A.~{Camurri}, G.~{Volpe}, N.~{Dael}, and K.~{Scherer},
	\newblock ``Technique for automatic emotion recognition by body gesture
	analysis,''
	\newblock in {\em Proc. of the IEEE Computer Society Conf. on Computer Vision
		and Pattern Recognition Workshops}, 2008.
	
	\bibitem{montepare1987gait}
	Joann~M. Montepare, Sabra~B. Goldstein, and Annmarie Clausen,
	\newblock ``The identification of emotions from gait information,''
	\newblock {\em Journal of Nonverbal Behavior}, vol. 11, no. 1, pp. 33--42,
	1987.
	
	\bibitem{li2016emotion}
	Shun Li, Liqing Cui, Changye Zhu, Baobin Li, Nan Zhao, and Tingshao Zhu,
	\newblock ``Emotion recognition using {Kinect} motion capture data of human
	gaits,''
	\newblock {\em PeerJ}, 2016.
	
	\bibitem{chen2017gif}
	W.~{Chen}, O.~O. {Rudovic}, and R.~W. {Picard},
	\newblock ``{GIFGIF+:} collecting emotional animated {GIFs} with clustered
	multi-task learning,''
	\newblock in {\em {Proc. of ACII}}, 2017, pp. 510--517.
	
	\bibitem{li2015cloud}
	N.~{Li}, Y.~{Xia}, and Y.~{Xia},
	\newblock ``Semi-supervised emotional classification of color images by
	learning from cloud,''
	\newblock in {\em Proc. of the Intl. Conf. on Affective Computing and
		Intelligent Interaction}, 2015, pp. 84--90.
	
	\bibitem{barros2018omg}
	P.~{Barros}, N.~{Churamani}, E.~{Lakomkin}, H.~{Siqueira}, A.~{Sutherland}, and
	S.~{Wermter},
	\newblock ``The {OMG-Emotion} behavior dataset,''
	\newblock in {\em Proc. of the Intl. Joint Conf. on Neural Networks}, 2018.
	
	\bibitem{baveye2015videoprediction}
	Y.~{Baveye}, E.~{Dellandrea}, C.~{Chamaret}, and L.~{Chen},
	\newblock ``Deep learning vs. kernel methods: Performance for emotion
	prediction in videos,''
	\newblock in {\em {Proc. of ACII}}, 2015, pp. 77--83.
	
	\bibitem{kollias2017recognition}
	Dimitrios Kollias, Mihalis~A Nicolaou, Irene Kotsia, Guoying Zhao, and Stefanos
	Zafeiriou,
	\newblock ``Recognition of affect in the wild using deep neural networks,''
	\newblock in {\em Proc. of the IEEE Conf. on Computer Vision and Pattern
		Recognition Workshops}, 2017, pp. 26--33.
	
	\bibitem{zafeiriou2017aff}
	Stefanos Zafeiriou, Dimitrios Kollias, Mihalis~A Nicolaou, Athanasios
	Papaioannou, Guoying Zhao, and Irene Kotsia,
	\newblock ``Aff-wild: Valence and arousal 'in-the-wild' challenge,''
	\newblock in {\em Proc. of the IEEE Conf. on Computer Vision and Pattern
		Recognition Workshops}, 2017, pp. 34--41.
	
	\bibitem{tzirakis2019real}
	Panagiotis Tzirakis, Stefanos Zafeiriou, and Bj{\"o}rn Schuller,
	\newblock ``Real-world automatic continuous affect recognition from audiovisual
	signals,''
	\newblock in {\em Multimodal Behavior Analysis in the Wild}. Elsevier, 2019.
	
	\bibitem{goodfellow2016deep}
	Ian Goodfellow, Yoshua Bengio, and Aaron Courville,
	\newblock {\em Deep learning},
	\newblock MIT press, 2016.
	
	\bibitem{krizhevsky2012imagenet}
	Alex Krizhevsky, Ilya Sutskever, and Geoffrey~E Hinton,
	\newblock ``Imagenet classification with deep convolutional neural networks,''
	\newblock in {\em Advances in neural information processing systems}, 2012, pp.
	1097--1105.
	
	\bibitem{young2018recent}
	Tom Young, Devamanyu Hazarika, Soujanya Poria, and Erik Cambria,
	\newblock ``Recent trends in deep learning based natural language processing,''
	\newblock {\em IEEE Computational Intelligence Magazine}, vol. 13, no. 3, 2018.
	
	\bibitem{karpathy2014video}
	Andrej Karpathy, George Toderici, Sanketh Shetty, Thomas Leung, Rahul
	Sukthankar, and Li~Fei-Fei,
	\newblock ``Large-scale video classification with convolutional neural
	networks,''
	\newblock in {\em Proc. of the IEEE Conf. on Computer Vision and Pattern
		Recognition}, 2014.
	
	\bibitem{ji2013convolutional}
	Shuiwang Ji, Wei Xu, Ming Yang, and Kai Yu,
	\newblock ``3d convolutional neural networks for human action recognition,''
	\newblock {\em IEEE Trans. on Pattern Analysis and Machine Intelligence}, vol.
	35, pp. 221–231, 2013.
	
	\bibitem{jia2014caffee}
	Yangqing {Jia}, Evan {Shelhamer}, Jeff {Donahue}, Sergey {Karayev}, Jonathan
	{Long}, Ross {Girshick}, Sergio {Guadarrama}, and Trevor {Darrell},
	\newblock ``{Caffe: Convolutional Architecture for Fast Feature Embedding},''
	\newblock {\em arXiv e-prints}, Jun 2014.
	
	\bibitem{yannakakis2012playermodeling}
	Georgios~N. Yannakakis, Pieter Spronck, Daniele Loiacono, and Elisabeth
	Andr\'{e},
	\newblock ``Player modeling,''
	\newblock in {\em Artificial and Computational Intelligence in Games (Dagstuhl
		Seminar 12191).}, pp. 45--59. 2012.
	
	\bibitem{pedersen2009experience}
	Chris Pedersen, Julian Togelius, and Georgios~N. Yannakakis,
	\newblock ``Modeling player experience in super mario bros,''
	\newblock in {\em Proc. of the Intl. Conf. on Computational Intelligence and
		Games}, 2009.
	
	\bibitem{shaker2013visual}
	Noor Shaker, Stylianos Asteriadis, Georgios~N. Yannakakis, and Kostas
	Karpouzis,
	\newblock ``Fusing visual and behavioral cues for modeling user experience in
	games,''
	\newblock {\em IEEE Trans. on System, Man and Cybernetics}, vol. 43, no. 6,
	2013.
	
	\bibitem{yannakakis2018artificial}
	Georgios~N. Yannakakis and Julian Togelius,
	\newblock {\em {Artificial Intelligence and Games}},
	\newblock Springer, 2018,
	\newblock \url{http://gameaibook.org}.
	
	\bibitem{guzdial2016deep}
	Matthew Guzdial, Nathan Sturtevant, and Boyang Li,
	\newblock ``Deep static and dynamic level analysis: A study on infinite
	mario,''
	\newblock in {\em Proc. of the AIIDE workshop on Experimental AI in Games},
	2016.
	
	\bibitem{martinez2013learning}
	Hector~P Martinez, Yoshua Bengio, and Georgios~N Yannakakis,
	\newblock ``Learning deep physiological models of affect,''
	\newblock {\em IEEE Computational Intelligence Magazine}, vol. 8, no. 2, pp.
	20--33, 2013.
	
	\bibitem{camilleri2017generalmodels}
	Elizabeth Camilleri, Georgios~N. Yannakakis, and Antonios Liapis,
	\newblock ``Towards general models of player affect,''
	\newblock in {\em Proc. of the Intl. Conf. on Affective Computing and
		Intelligent Interaction}, 2017.
	
	\bibitem{schindler2008action}
	Konrad Schindler and Luc~J Van~Gool,
	\newblock ``Action snippets: How many frames does human action recognition
	require?,''
	\newblock in {\em Proc. of the IEEE Intl. Conf. on Computer Vision and Pattern
		Recognition}, 2008.
	
	\bibitem{makantasis2016deep}
	Konstantinos Makantasis, Anastasios Doulamis, Nikolaos Doulamis, and
	Konstantinos Psychas,
	\newblock ``Deep learning based human behavior recognition in industrial
	workflows,''
	\newblock in {\em Proc. of the Intl. Conf. on Image Processing}. IEEE, 2016,
	pp. 1609--1613.
	
	\bibitem{yannakakis2018ordinal}
	Georgios~N Yannakakis, Roddy Cowie, and Carlos Busso,
	\newblock ``{The Ordinal Nature of Emotions: An emerging approach},''
	\newblock {\em {IEEE Trans. on Affective Computing}}, 2018.
	
	\bibitem{martinez2014don}
	Hector~P Martinez, Georgios~N Yannakakis, and John Hallam,
	\newblock ``{Don't classify ratings of affect; rank them!},''
	\newblock {\em IEEE Trans. on Affective Computing}, vol. 5, no. 3, pp.
	314--326, 2014.
	
	\bibitem{selvaraju2017grad}
	Ramprasaath~R Selvaraju, Michael Cogswell, Abhishek Das, Ramakrishna Vedantam,
	Devi Parikh, and Dhruv Batra,
	\newblock ``Grad-cam: Visual explanations from deep networks via gradient-based
	localization,''
	\newblock in {\em Proc. of the IEEE Intl. Conf. on Computer Vision}, 2017.
	
\end{thebibliography}

\end{document}